\documentclass[twocolumn]{aastex63}
\pdfoutput=1 

\usepackage{tabularx}

\received{}
\revised{}
\accepted{}

\shortauthors{Soraisam et al.}

\begin{document}

\title{AT~2020iko: a WZ Sge-type DN candidate with an anomalous precursor event}

\correspondingauthor{Monika Soraisam}
\email{soraisam@illinois.edu}

\author[0000-0001-6360-992X]{Monika D.~Soraisam}
\affiliation{National Center for Supercomputing Applications, University of Illinois at Urbana-Champaign, Urbana, IL 61801, USA}
\affiliation{Department of Astronomy, University of Illinois at Urbana-Champaign, Urbana, IL 61801, USA}

\author{Sarah~R.~DeSantis}
\affiliation{Steward Observatory, University of Arizona, 933 North Cherry Avenue, Tucson, AZ 85721, USA}

\author[0000-0003-1700-5740]{Chien-Hsiu Lee}
\affiliation{NSF's National Optical-Infrared Astronomy Research Laboratory, 
950 North Cherry Avenue, Tucson, AZ 85719, USA}

\author[0000-0001-6685-0479]{Thomas Matheson}
\affiliation{NSF's National Optical-Infrared Astronomy Research Laboratory, 950 North Cherry Avenue, Tucson, AZ 85719, USA}

\author[0000-0001-6022-0484]{Gautham Narayan}
\affiliation{Department of Astronomy, University of Illinois at Urbana-Champaign, 
Urbana, IL 61801, USA}

\author[0000-0002-6839-4881]{Abhijit Saha}
\affiliation{NSF's National Optical-Infrared Astronomy Research Laboratory, 950 North Cherry Avenue, Tucson, AZ 85719, USA}

\author[0000-0003-4102-380X]{David~J.~Sand}
\affiliation{Steward Observatory, University of Arizona, 933 North Cherry Avenue, Tucson, AZ 85721, USA}

\author[0000-0002-2744-714X]{Carl Stubens}
\affiliation{NSF's National Optical-Infrared Astronomy Research Laboratory, 950 North Cherry Avenue, Tucson, AZ 85719, USA}

\author{Paula Szkody}
\affiliation{University of Washington, Department of Astronomy, Box 351580, Seattle, WA 98195, USA}

\author[0000-0002-9508-1629]{Nicholas Wolf}
\affiliation{NSF's National Optical-Infrared Astronomy Research Laboratory, 950 North Cherry Avenue, Tucson, AZ 85719, USA}

\author{Samuel~D.~Wyatt}
\affiliation{Steward Observatory, University of Arizona, 933 North Cherry Avenue, Tucson, AZ 85721, USA}

\author{Ryohei Hosokawa}
\affiliation{Tokyo Institute of Technology, 2 Chome-12-1 Ookayama, Meguro City, Tokyo 152-8550, Japan}

\author{Nobuyuki Kawai}
\affiliation{Tokyo Institute of Technology, 2 Chome-12-1 Ookayama, Meguro City, Tokyo 152-8550, Japan}

\author{Katsuhiro~L.~Murata}
\affiliation{Tokyo Institute of Technology, 2 Chome-12-1 Ookayama, Meguro City, Tokyo 152-8550, Japan}

\begin{abstract}

The ongoing Zwicky Transient Facility (ZTF) survey is generating a massive alert rate from a variety of optical transients and variable stars, which are being filtered down to subsets meeting user-specified criteria by broker systems such as ANTARES. In a beta implementation of the algorithm of Soraisam et al.\ (2020) on ANTARES, we flagged AT~2020iko from the ZTF real-time alert stream as an anomalous source. This source is located close to a red extended SDSS source. In the first few epochs of detection, it exhibited a V-shaped brightness profile, preceded by non-detections both in ZTF and in ASAS-SN extending to 2014. Its full light curve shows a precursor event, followed by a main superoutburst and at least two rebrightenings. A low-resolution spectrum of this source points to a dwarf nova (DN) nature. Although some of the features of AT~2020iko indicate an SU~UMa-type DN, its large amplitude, presence of rebrightenings, and inferred supercycle period of $\geq6$~yr are in favor of AT~2020iko being a new WZ~Sge-type dwarf nova candidate, a subset of rare DNe consisting of extreme mass-ratio ($<0.1$) binaries with orbital period around the period minimum. AT~2020iko's precusor event brightened by 6.5~mag, while its decay spanned 3-5 mag. We speculate this superoutburst is associated with a less expanded accretion disk than in typical superoutbursts in WZ~Sge systems, with the large depth of the precursor decay implying an extremely small mass-ratio. To the best of our knowledge, such a precursor event has not been recorded for any DN. This result serves to demonstrate the efficacy of our real-time anomaly search algorithm. 
\end{abstract}


\section{Introduction} \label{sec:intro}
Current wide-field sky surveys are boosting the density of, and even occasionally populating, the parameter space of observed astrophysical sources. The Zwicky Transient Facility (ZTF; \citealt{Bellm-2019}) is a leader in 
optical time-domain astronomy. ZTF is surveying the entire visible northern sky using a CCD camera with a 47~${\rm deg}^2$ field of view fitted to the 48~inch Samuel Oschin Schmidt Telescope at the Palomar Observatory. `Alerts', i.e., detections of flux- or position-varying sources at ${\rm SNR}\geq5$ \citep{Masci-2019}, from the public part of ZTF (comprising 40\% of observing time) are being streamed to event-brokers \citep{Patterson-2019}. The latter are software systems that add values to alerts, custom-filter them to a manageable level, and deliver them to the users near real-time; examples include ANTARES (\citealt{Saha-2016}, Matheson et al.\ 2020 submitted), AMPEL \citep{AMPEL}, LASAIR \citep{LASAIR}, ALeRCE \citep{Alerce}, and MARS{\footnote{https://mars.lco.global}}.  

An important goal of the real-time delivery of (filtered) alerts is to facilitate discovery and characterization of the nature, via follow-up observations, of relatively rare (e.g., in terms of rate) short-lived sources. Algorithms aimed toward such discoveries for astronomical time-series data have been developed \citep[e.g.,][]{Giles-2019, Ishida-2019, Soraisam-2020}. In this paper, we present the discovery and multi-wavelength follow-up results of a new WZ~Sge-type dwarf nova (DN) candidate, from a beta implementation of such an algorithm in the real-time alert environment.

Dwarf novae are a subclass of cataclysmic variables (CVs) comprising an accreting white dwarf (WD) in a binary system with typically a late-type main-sequence companion, whereby the WD undergoes semi-regular outbursts \citep{Warner-2003}. The most widely studied theoretical model for the DN outbursts invokes a thermal instability in the accretion disk, which acts to increase the viscosity (parameterized by $\alpha$; \citealt{Shakura-1973}) resulting in a runaway increase of the disk temperature and mass accretion onto the WD, which, in turn, transforms the accretion disk to a cool state, and the cycle repeats \citep[e.g.,][]{Smack-1971, Meyer-1981, Osaki-1996, Lasota-2001, Hameury-2019}. 
The DN outbursts are a common phenomenon, accounting for the majority of the observed optical transients or stochastic variable stars in the Galaxy in current synoptic time-domain surveys \citep[e.g.,][]{Szkody-2020}. These objects have provided significant insights into the properties of accreting WDs, in particular their accretion disks, and the evolution and population of low-mass stellar binaries \citep{Pala-2020}. 

Photometrically, DN outbursts are often characterized by an amplitude of several magnitudes ($\sim 2\mbox{--}5$~mag) lasting for a few days and recurring on timescales of months. However, there is a rarer subclass called WZ~Sge-type DNe, which show a larger-amplitude (reaching up to $\sim 9$~mag) superoutburst lasting for week(s), superposed with periodic fluctuations (called superhumps), multiple rebrightenings, and a long recurrence period on the order of a decade or more \citep{Kato-2015}. They are at the extreme tail of the SU~UMa-type DNe, which show both normal and superoutbursts recurring on a timescale of typically a few months. For the following, the reader should keep in mind this subclass relation whenever aspects of either class of systems are discussed.

WZ Sge-type systems have a short orbital period, close to the period minimum ($\sim 80$~min), a very low value of the viscosity parameter during quiescence $\alpha_{\rm cold}$, and a highly asymmetric mass-ratio with the donor star mass less than a tenth of that of the WD (\citealt{Smak-1993, Osaki-2002} and references therein). Their early-phase (within the first ten days or so of the outburst) superhumps correlate with the orbital period of the system -- the inclination-dependent photometric manifestation of tidal dissipation of angular momentum resulting from the 2:1 resonance between the accretion disk and the orbital motion in close binaries with extreme mass-ratios \citep[e.g.,][]{Lin-1979}. Their later-phase superhumps, on the other hand, have a period that is slightly longer than the orbital period, and are related to the precession of the tidally deformed eccentric disk excited at the 3:1 resonance radius, common across  SU~UMa-type DNe \citep{Whitehurst-1988}. 

Although the number of WZ Sge-type DNe detected has significantly increased with the advent of wide-field surveys, they remain relatively rare, with their discovery rate being a few tens per year \citep{Kato-2015}. For comparison, there are, for example, a few hundred type Ia supernovae detected per year. This is likely due to the fact that dedicated SN searches covering large areas with good enough cadence are quite popular, while there are no dedicated searches with proper cadences for such DNe. 
Next-generation wide-field surveys, specifically the Rubin Observatory's Legacy Survey of Space and Time (LSST; \citealt{Ivezic-2019}), have the potential to discover many such transients, particularly opening up the faint-magnitude space. However, its main survey component, namely the Wide-Fast-Deep survey, which is currently baselined to a sampling cadence of around 10-20 days per passband (\citealt{Marshall-2017}, Chapter 2) is not ideal to detect and characterize fast-evolving outbursts from CVs. Specialized mini-surveys focusing on the Galactic Plane or a few nearby galaxies with a much higher cadence (e.g., deep-drilling-like; \citealt{Marshall-2017}, Chapter 4), which have been proposed, would be more favorable to this end.

This paper is organized as follows. Section~\ref{sec:TS} outlines the discovery of our source, while the follow-up observations are detailed in Section~\ref{sec:fup}. We discuss the observation results in Section~\ref{sec:dis} and conclude with Section~\ref{sec:con}.

\begin{figure}
    \centering
    \includegraphics[width=80mm]{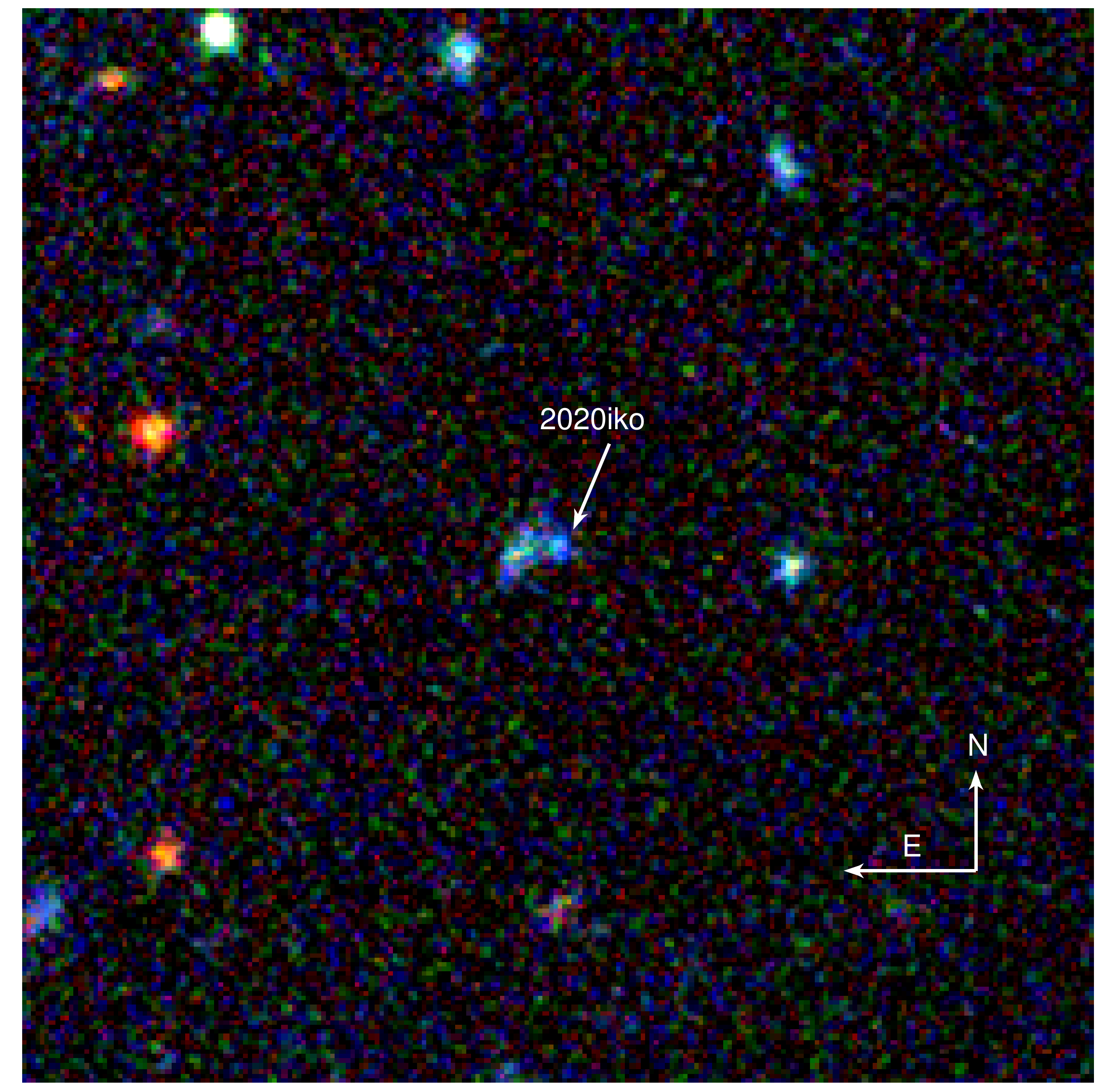}
    \caption{Color cutout image ($1'\times1'$) of 2020iko (indicated by the white arrow) from the Legacy Surveys. An extended source is also visible immediately to the east (i.e., left) of 2020iko. 
    }
    \label{fig:cutout}
\end{figure}

\begin{figure*}
    \centering
    \includegraphics[width=170mm]{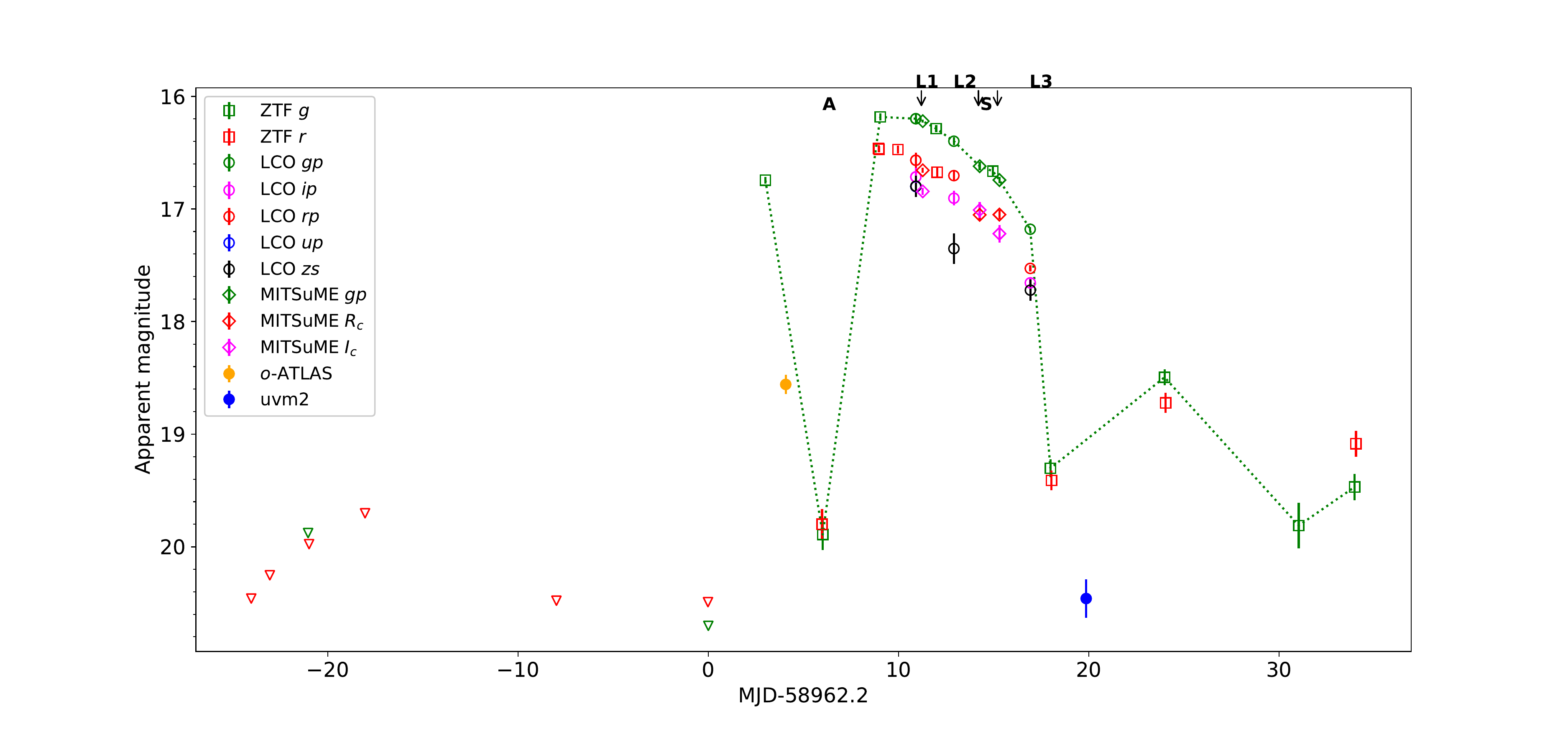}\\
    \includegraphics[width=170mm]{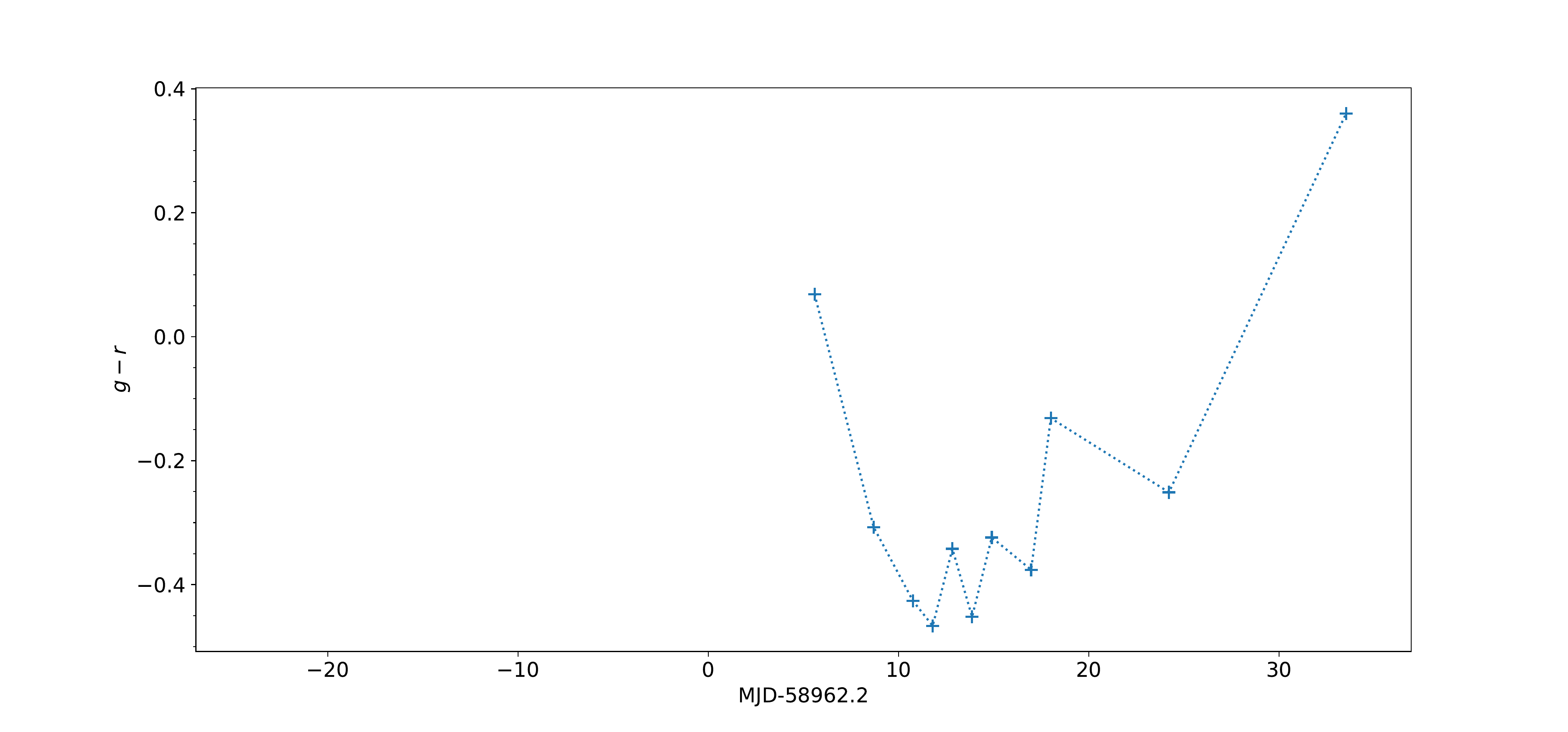}
    \caption{{\it Top}: Light curve of 2020iko aggregating the measurements from ZTF, ATLAS, LCO, MITSuME and {\it Swift}, as indicated by the legend. The LCO observations from the three nights, indicated by L1, L2 and L3, as well as the MITSuME observations indicated by the arrows are shown zoomed-in in Fig.~\ref{fig:days}. Triangles denote the upper limits from ZTF in $g$ (green) and $r$ (red) passbands. `A' marks the epoch when the source was flagged by our anomaly filter, and S, the epoch of spectroscopic observation (Sect.~\ref{spec}). The time axis is shown relative to the epoch of latest ZTF upper limit. The green dotted line connects the green data points and is meant to guide the eye regarding the evolution of the outburst. Note that the error bars are much smaller than the symbols for many of the data points.  {\it Bottom}: Color ($g-r$) evolution of 2020iko. }
    \label{fig:lc}
\end{figure*}

\begin{figure*}[t]
    \centering
    \includegraphics[width=170mm]{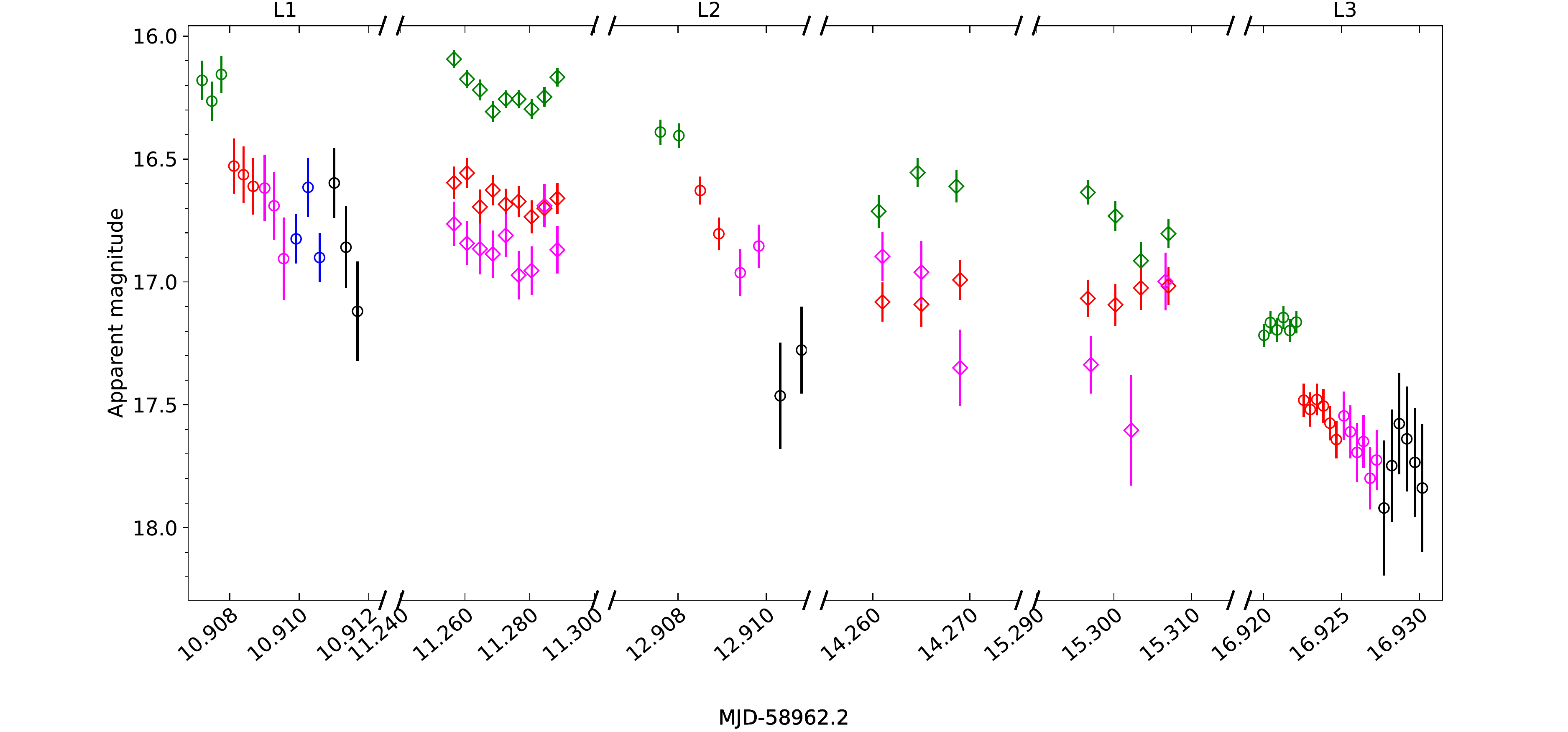}
    \caption{
    Light curve of 2020iko from follow-up LCO observations on the three epochs L1, L2, and L3, and MITSuME observations indicated by the arrows in Fig.~\ref{fig:lc}. The corresponding passbands of the data points are the same as shown in the legend of Fig.~\ref{fig:lc}. There were no $up$ measurements on the last two LCO epochs, L2 and L3.
    }
    \label{fig:days}
\end{figure*}

\section{Anomaly Transient Selection} \label{sec:TS}
We have implemented a real-time filter on ANTARES based on the algorithm of \citet{Soraisam-2020} to flag peculiar astrophysical sources from the ZTF alert stream. To this end, we use a sample of tens of thousands of persistent variable sources from ZTF, whose labels are obtained from cross-matching with external catalogs in the ANTARES processing. The relevant catalogs include those of ASAS-SN variable stars \citep{Jayasinghe-2018} and quasars and AGNs \citep{Veron-2010}. We use the ZTF light curves of these variable sources to construct the conditional probabilities of magnitude changes and pseudo-colors for given time intervals. These distributions are used to evaluate the likelihood score of consistency with this population of variable sources for any newly detected alert (see \citealt{Soraisam-2020} for details).

ZTF20aawbodq (AT2020iko, hereafter 2020iko) was flagged with one of the lowest likelihood scores, thus denoting a peculiar candidate \citep{ATel}. The source is located at (J2000) RA=09:03:26.23 DEC=+47:31:59.14 ($l=172.12759$, $b=41.57730$), within $0.5''$ of the extended SDSS source J0903+4731 \citep{Alam-2015}. This appears in line with the ZTF {\it r}-band nearest neighbor from the deep reference image (used in difference imaging) at a distance of $1.83''$ (less than the typical seeing of $2''$ for ZTF) from 2020iko, a magnitude of 22.89, and a star-galaxy score of 0.4 indicating extendedness{\footnote{The ZTF alerts data for 2020iko are accessible at \url{https://antares.noirlab.edu/loci/ANT2020ea3ig}}}. There is no {\it g}-band nearest neighbor in the ZTF data within $2''$. Given the close proximity to the extended source SDSSJ0903+4731, we initially suspected 2020iko to be an extragalactic transient.

Examining deeper images along with their source catalog from the recent eighth data release of the DESI Legacy Imaging Surveys \citep{Dey-2019}, we find two sources around the position of 2020iko, for which a cutout of the color ($g,r,z$) image stack is shown in Fig.~\ref{fig:cutout}. Out of the two, one is a point-like source at a distance of $0.71''$ from the ZTF location of 2020iko with magnitudes $g=23.27, r=23.27, z=23.51$, and the other is an extended source at a distance of $1.75''$ with magnitudes $g=23.0, r=22.26, z=21.85$. Thus, the closer point source, which is marked with the arrow in Fig,~\ref{fig:cutout} and is slightly fainter than the extended source, is likely the progenitor of 2020iko. 

The aggregated light curve of 2020iko is shown in Fig.~\ref{fig:lc}, which includes data from ZTF, ATLAS \citep{Tonry-2018}, as well as follow-up observations (see Sect.~\ref{sec:fup}). 
Table~\ref{tab:lc_data} lists all the photometric measurements. 
Our filter first flagged this source around MJD 58968 corresponding to the sharp decline of $\sim3$~mag in 3~d. The public ZTF survey first detected it on 26 April 2020 UT (MJD 58965), at $g=16.74$, preceded by non-detections, latest of which was on 23 April 2020 UT (MJD 58962) used as the time offset in Fig.~\ref{fig:lc}. Overall, the complete light curve seems to exhibit three features -- a precursor decline, a main outburst with a plateau-like profile, and post-outburst rebrightenings -- reminiscent of DNe as discussed above.

\section{Follow-up} \label{sec:fup}
We initiated optical photometric and spectroscopic follow-up observations with Las Cumbres Observatory (LCO; \citealt{Brown-2013}) and MITSuME \citep{Kotani-2005} facilities, as well as high-energy observations with {\it Swift} \citep{Gehrels-2004}.

\subsection{LCO photometry} \label{LCO}
We performed multi-wavelength follow-up observations using the Sinistro imager on the LCO 1-m telescope at the McDonald Observatory on three nights -- UT 4 May (MJD 58973), 6 May (MJD 58975; TOM2020A-012; PI:~Lee), and 10 May (MJD 58979; TOM2020A-011; PI:~Narayan), labelled as L1, L2 and L3, respectively, in Fig.~\ref{fig:lc}. The exposure times for the different passbands ($gp, rp, ip, zs$){\footnote{The $up, gp, rp, ip, zs$ filters on the LCO Spectral camera correspond to those of SDSS/Pan-STARRS }} per image are (3s, 3s, 3s, 10s) on epoch L1, and (10s, 10s, 10s, 15s) on epochs L2 and L3. We also took a sequence of $up$ exposures on epoch L1, each of 10s. We collected a total of 3, 2 and 6 exposures per passband on L1, L2 and L3, respectively.

We obtained the BANZAI pipeline-reduced images from LCO \citep{McCully-2018} and used them to perform forced aperture photometry at the position of 2020iko. We calibrated the photometric results using PS1 measurements of sources in the same field (typically a few dozen), and SDSS measurements for the $up$-band. The zero point errors are only a few millimags. The error-weighted average per passband for a given epoch is shown in the aggregated light curve (Fig.~\ref{fig:lc}). As evident from the plot, the LCO photometry data (particularly $gp$ and $rp$) smoothly follow the plateau-like profile giving credence to our calibration.   

The individual LCO measurements on each of these three nights are also shown expanded in Fig.~\ref{fig:days}. As can be seen, the redder data points (particularly $zs$) are marred by large errors. However, a hint of coherent intra-night variability can be seen mainly for $rp$ and $ip$ measurements (e.g., on epoch L3) at the level of 0.2~mag.

\subsection{MITSuME photometry}\label{MITS}
2020iko was also observed simultaneously in three passbands ($gp, R_{c}, I_{c}$) with the MITSuME 0.5-m telescope at Akeno Observatory on three nights -- UT 4 May (MJD 58973), 7 May (MJD 58976), and 8 May (MJD 58977), marked by arrows in Fig.~\ref{fig:lc}. A sequence of 60s exposures were taken on each of these nights. We performed photometry for these observations in a similar manner to that of LCO data, and the error-weighted average per passband for each night is shown in (Fig.~\ref{fig:lc}). 

For the MITSuME observations, we have a larger number of measurements per night than with LCO. We therefore bin them into 5~min bins to reduce the measurement noise and the results are shown in Fig.~\ref{fig:days}. A clear signature of intra-night variability at a level of 0.2~mag (for $gp$) can be seen for the 4 May (MJD 58973) MITSuME observations with the longest intra-night baseline of around 50~min. This interval is, however, smaller than the minimum orbital period of CV systems and therefore it is not sufficient to perform period analysis, which requires covering at least 1.5 cycles for any meaningful inference. Nevertheless, the variability level matches the typical amplitude of superhump variability (Sect.~\ref{sec:dis}).

\subsection{Optical color evolution}
We also compute the color evolution using $g$/$gp$ and $r$/$rp$/$R_{c}$ data points. We bin the light curve into 1~day bins, and take the mean magnitude in a given passband if there is more than one measurement. 
It is to be noted that the photometric measurements of ZTF are obtained from difference images, where the extended red source in the background has been subtracted (assuming it does not vary). For the follow-up observations, the photometry is performed without image subtraction. However, the extended source is very faint ($r\sim 22.3$~mag), and therefore it is not expected to significantly influence the color data points computed using those measurements, which were all taken around the bright plateau of the light curve. 
The result is shown in the bottom panel of Fig.~\ref{fig:lc}. 
The color has been corrected for foreground extinction using the reddening value of $E(B-V)=0.026$ in the farthest-distance bin from the 3D reddening map of \citet{Green-2018}.  

As can be seen, the source is quite red with $g-r\sim0.1$ at the precursor minimum, but morphs to a blue color with $g-r\sim-0.3$ three days later at the observed peak of the main outburst. The emission continues to grow bluer in time during the gradually-declining plateau in the light curve reaching $g-r\sim-0.45$, later fluctuating slightly at a level of around 0.1~mag, possibly related to incomplete phase coverage for the follow-up observations. This is immediately followed by a color inversion as the light curve drops sharply in brightness around MJD 58980, but later grows bluer again during the subsequent first rebrightening. The light curve seems to indicate a second rebrightening peak beyond the last epoch of observation by ZTF on  MJD 58996 (see Sect.~\ref{sec:dis}), where the color is the reddest ($g-r\sim0.35$). Thus, the color trend of 2020iko can be summarized as bluer when bright and redder when faint.       

During a DN outburst, the accretion disk is optically thick and its emission can thus be modeled with a blackbody with radially varying temperature profile \citep{Horne-1985}. 
Assuming the optical emission to originate from the same outermost region, we can make a crude estimate of the evolution of the color temperature using Fig.~\ref{fig:lc} (bottom panel). Noting that the passbands $g$ and $r$ do not have much constraining power at high temperatures when they are located at the Rayleigh-Jeans tail, we obtain a color temperature $>10^{4}$~K at $g-r\lesssim0.1$, while at the reddest point $g-r\sim0.35$ during the rebrightening phase, the color temperature is around 7500~K.

The overall color trend of 2020iko appears to be similar to observations of superoutbursts of other DN systems. For example, the 2001 superoutburst of the prototype WZ~Sge with $UBVRI$ observations showed a blue color during its initial superoutburst phase and a redder color during its decline, which later transitioned to a blue color again at its post-superoutburst rebrightening peak, though not as blue as during the superoutburst \citep[see][]{Howell-2004}. Based on the observed color temperatures, \citet{Howell-2004} characterized the observed rebrightening of WZ Sge as a cyclical transition of the accretion disk from a neutral state to an ionized state.  
A similar behavior is also shown by the optical color,  in particular, $V-R_{c}$, of another WZ~Sge-type DN SDSS~J102146.44+234926.3 \citep{Uemura-2008}.

\begin{figure}
    \centering
    \includegraphics[width=80mm]{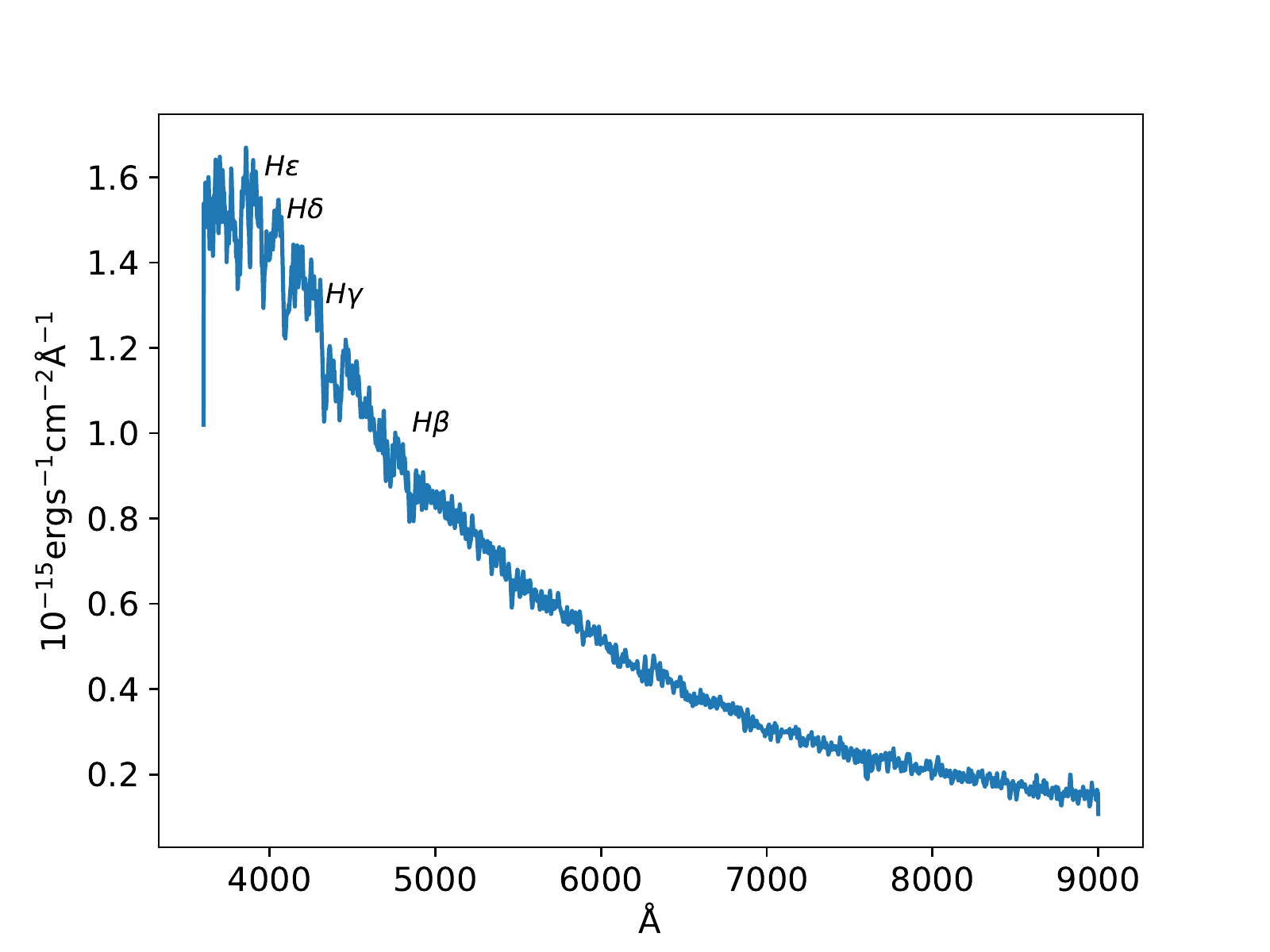}
    \caption{Optical spectrum of 2020iko observed with the LCO FLOYDS spectrograph. The spectrum has been smoothed using a running mean box kernel with width 3 pixels.
    The Balmer absorption lines are marked in the plot.}
    \label{fig:spec}
\end{figure}

\subsection{LCO spectroscopy} \label{spec}
We also obtained a spectrum of the source on 7 May 2020 UT (MJD 58976), with the  cross-dispersed low-resolution FLOYDS spectrograph of LCO, mounted on its 2-m telescope on the Haleakala Observatory (LCO2020A-006; PI:~Sand). We used a slit width of $2''$ with an exposure time of 2700~sec. As can be seen from Fig.~\ref{fig:lc}, 2020iko was brighter than 17~mag when the spectrum was taken. Hence, despite the large slit width, the observation is not expected to be affected by the faint ($\gtrsim 22$~mag) nearby extended source (see Sect.~\ref{sec:TS}). LCO performs the data reduction automatically based on the pipeline of \citet{Valenti-2014}. The extracted wavelength- and flux-calibrated 1D spectrum of 2020iko delivered by the LCO pipeline is shown in Fig.~\ref{fig:spec}.

The source exhibits a blue continuum. Though the spectrum is quite noisy, distinct absorption features of higher-order Balmer lines, in particular $H\beta$ through $H\epsilon$, can be seen at $z=0$, which are marked in the plot; no $H\alpha$ feature is, however, visible in the spectrum. There is an absorption feature around 4425~\AA, near $H\gamma$, which we verified as likely an artifact due to the imperfect flattening in the blue side and not due to Ca I. The spectrum is, thus, typical of a DN during outburst, wherein the accretion disk transitions to a hot, optically-thick state leading to the blue continuum and its absorption by the surface layers resulting in the Balmer absorption lines \citep[e.g.,][]{Clarke-1984, Szkody-2002, Szkody-2003}.

In addition to the blue continuum and Balmer absorption lines, emission lines of He~II and C III/N III have been detected during the early outburst stage in WZ~Sge type DNe with time-resolved spectra  available \citep[see][and references therein]{Kato-2015}. Additionally, Na~I~D absorption line has been observed during the early stage in the prototype WZ~Sge \citep{Nogami-2004} and during the rebrightening phase in EG~Cnc \citep{Patterson-1998}. These lines are not detected in the spectrum of 2020iko. However, its spectrum was obtained only during the late phase of its gradually-declining plateau (Fig.~\ref{fig:lc}), which may explain the absence of these spectral features. The equivalent widths (EWs) of the higher-order Balmer absorption lines of 2020iko are around 5~\AA\, each, slightly larger than the EWs of 2-3~\AA\, for these lines observed around a similar stage for the superoutburst of the prototype WZ~Sge \citep{Nogami-2004}. However, we consider them to be broadly consistent given the lower signal-to-noise ratio of our spectrum and more importantly the poor spectral resolution of FLOYDS (ranging from 8.5~\AA\, on the blue side to 17~\AA\, on the red side), which make it difficult to determine precise EWs for 2020iko.

\subsection{{\it Swift} observations} \label{xray}
We also requested a {\it Swift} ToO observation with XRT and UVOT, which were executed about a week later on 13 May 2020 UT (MJD 58982) around the minimum of the decline of the main outburst. We analyzed the data using the {\it Swift} software package from HEASOFT. No X-ray emission was detected in an exposure of around 0.5~ks, and we obtained the $3\sigma$ upper limit on the count rate of $0.0267~{\rm{counts~s^{-1}}}$ by running \texttt{sosta}. The source was, however, detected with the UVOT/uvm2. We used \texttt{uvotsource} to measure its magnitude, which is indicated by the blue filled circle in Fig.~\ref{fig:lc}. 
Apart from the fact that the source exhibited UV emission, the limited number of uvm2 as well as $up$ measurements are unconstraining, nevertheless we show them in the light curve for completeness.

\section{Discussion} \label{sec:dis}
As mentioned in Sect.~\ref{sec:intro}, one of the distinctive features for WZ Sge-type DNe (and generally SU~UMa systems) is the presence of superhumps with amplitudes around $0.2\mbox{--}0.3$~mag during the superoutbursts. Their detection requires extensive observations with minute-scale cadence, which, unfortunately, were not obtained for 2020iko. However, the few intra-night observations (Fig.~\ref{fig:days}) indicate a similar level of variability at around 0.2~mag.

Based on the counterpart for 2020iko in the deep Legacy image (Sect.~\ref{sec:TS}), we obtain an amplitude of 7.1~mag in the $g$-band and 6.8 in the $r$-band for its superoutburst. Its precursor outburst (Sect.~\ref{sec:precursor}) stands out with an amplitude of 6.5~mag in the $g$-band. The duration of the superoutburst plateau is approximately 10~days, which is somewhat shorter than the typical duration of tens of days for WZ~Sge-type DNe. Also, almost all such DNe show only superoutbursts with no precursors, which are, on the other hand, common for SU~UMa-type systems. In WZ Sge systems, the outer edge of the accretion disk reaches the 2:1 resonance radius at the onset of a typical superoutburst. This radius serves as a tidal wall, where matter is dammed up due to very effective tidal removal of angular momentum from the disk \citep{Osaki-2003}. Additionally, with the large mass accumulated during their long quiescence, the surface density at the tidal wall is well above the critical value (below which no hot state of the disk is possible) and the disk remains in the hot (outburst) state for a long duration. There is, thus, no time for a cooling wave to propagate to result in a precursor event in such systems. 
However, there are exceptions such as AL~Com. This well-studied WZ~Sge-type DN exhibited a superoutburst in 2015 that shares many common characteristics with 2020iko. Its plateau lasted $\approx 10$~days and was preceded by a precursor (though not well separated; it is claimed to be the first precursor observed in any WZ~Sge type DNe) and showed a long post-superoutburst rebrightening \citep{Kimura-2016}. 
This particular superoutburst of AL~Com was attributed to limited expansion of the accretion disk, barely exceeding the 3:1 resonance radius and not reaching the 2:1 resonance radius at all.

Rebrightenings are typical in WZ~Sge-type DNe, even though their underlying physical mechanism is not yet well understood. In fact, the light curve of 2020iko hints at a second rebrightening (Fig.~\ref{fig:lc}). The last ZTF alert from it was on 27 May 2020 UT (MJD 58996). By this time, the source was setting -- it was already at an elevation $<40^{\circ}$. Hence, it is most likely that the 2020iko field was not observed on subsequent nights, which prevented capturing additional rebrightening events.      

Furthermore, to probe any possible historical outburst of 2020iko prior to ZTF, we queried the ASAS-SN database{\footnote{https://asas-sn.osu.edu/}} \citep{Shappee-2014, Kochanek-2017} for its light curve since the start of the ASAS-SN survey. The baseline of ASAS-SN extends to 2014 with an average cadence of 2~days, covering a typical depth of $V\sim17$~mag, which should be sufficient to detect at least the bright plateau phase of an outburst similar to that in Fig.~\ref{fig:lc}. However, no record of any historical outburst exists.

Taking all the above factors into account, we favor 2020iko to be a WZ~Sge-type DN candidate with a superoutburst and a precursor (see below). Its superoutburst is, however, different from the typical superoutbursts in other WZ~Sge type systems, which are associated with the extension of the accretion disk beyond the 2:1 resonance radius (see \citealt{Osaki-2002}). Instead, the underlying process in 2020iko's superoutburst is likely similar to that of AL~Com described above.

\subsection{Significance of 2020iko's precursor}\label{sec:precursor}
Some WZ~Sge systems have been observed with a dip in brightness in the middle of their superoutburst, such as ASASSN-15jd \citep{Kimura-2016b} and SSS~J122221.7-311523 \citep{Kato-2013a}. The duration of the pre-dip segment of the pleateau for these systems is typically 10-20~days. The dip in these systems is attributed to the slow growth of the 3:1 resonance tidal instability following the end of the 2:1 resonance instability, which operates during the first segment. A cooling wave is, thus, able to propagate in the disk in the meantime, leading to a drop in the brightness. If we were to assume 2020iko to have a large dip in the middle of its superoutburst instead of a precursor, the duration of its first plateau segment would be at most 3-4~days (see Fig.~\ref{fig:lc}). To explain such a significantly shorter first segment would require some additional mechanism that can accelerate the removal of matter from the 2:1 resonance region. In the absence of such a mechanism, we instead consider this event in 2020iko to be a precursor outburst.

In the thermal tidal instability (TTI) model \citep{Osaki-1996, Osaki-2003}, a superoutburst in SU~UMa type systems is triggered by a normal outburst. 
Depending on the final extension of the accretion disk during the precursor outburst and the growth rate of the resulting tidal instability, different time delays between the precursor and superoutburst are achieved.   
Many observed SU~UMa type systems show well-separated (i.e., 5-10~days) precursors to superoutbursts (for example, see \citealt{Mroz-2015} based on OGLE observations, \citealt{Osaki-2014} for V1504~Cyg and V344~Lyr, and \citealt{Barclay-2012} for NIK~1, the latter two works based on {\it Kepler} data). Those precursors typically exhibit a decay of 1-1.5~mag.{\footnote{One larger decay of 3~mag has been observed for V1504~Cyg with {\it Kepler} \citep[Figs.~1 and 3 of][]{Osaki-2014}, however these observations were heavily affected by noise, as described by \citet{Osaki-2014}.}} 
On the other hand, precursors with much shallower decay (around 0.5~mag or less) appear merged with their superoutbursts, as shown by many SU~UMa type systems including TW~Vir, SDSS~111236.7+002807 \citep{Dai-2016},  CRTS~J035905.9+175034 \citep{Littlefield-2018}, and also some superoutbursts of V1504~Cyg and V344~Lyr \citep{Wood-2011, Osaki-2013, Osaki-2014} (see also the theoretical models of \citealt{Howell-1995}). Even the precursor to the 2015 superoutburst of AL~Com (Sect.~\ref{sec:dis}) belongs to this category. Hereafter we refer to the former type as `normal precursors' and the latter as `merged precursors'. The normal precursors are identical in amplitude to the normal outbursts{\footnote{Statistical studies of properties of CVs, for example, as conducted by \citet{Drake-2014} show the distribution of outburst amplitudes extending to 8~mag. Though the CVs were not classified into sub-types, their orbital period distribution is concentrated at short orbital periods below the period gap. Thus, the CV outbursts in their sample are particularly biased toward superoutbursts, and hence the large amplitudes.}} in the corresponding systems \citep{Osaki-2003, Osaki-2014}, while the merged precursors exhibit a larger amplitude, though still $\lesssim6$~mag.

In the high-cadence {\it Kepler} observations of V1504~Cyg and V344~Lyr, \citet{Osaki-2014} found that superhumps already appeared in the decay phase of their normal precursor outbursts, which then grew in amplitude with the development of the main outbursts. This is evidence in support of tidal instability powering these superoutbursts. The authors also found some normal outbursts with the same behavior, but with aborted superhumps \citep{Osaki-2003}, which disappeared near quiescence, failing to excite a superoutburst.

As can be seen from Fig.~\ref{fig:lc}, 2020iko's precursor has been caught in its decline -- fading by 3~mag in 3~days. Following a 3~day gap, where there were no observations, the light curve shot up from its observed precursor minimum ($g\sim 20$) to the peak of its main outburst ($g\sim16$), while the separation between the two observed peaks (i.e., of the precursor and superoutburst) is 6~days. Clearly it is not a merged precursor, but neither is it a normal precursor (see below). With a decline rate of $1~{\rm mag} ~{\rm day}^{-1}$ for the precursor phase, it is unlikely this segment of the light curve faded to its progenitor magnitude of $23.27$ (Sect.~\ref{sec:TS}). Nevertheless, since the typical rise time for a superoutburst in DNe is around a day, it is possible for the precursor to have reached $22$~mag, which would imply a deep decay of $\sim 5$~mag.

Despite the similarity in the light curve morphology of 2020iko with SU~UMa type DN superoutbursts that show precursors, there are three main differences: 
\begin{itemize}
\item 2020iko has no recorded historical (normal or super) outbursts in $>6$~yr. This points to a very low value of the viscosity parameter $\alpha_{\rm cold}$ for the cold accretion disk during quiescence, characteristic of WZ~Sge systems \citep[e.g.,][]{Smak-1993, low-alpha}. 
\item Its precursor has an amplitude of 6.5~mag (considering the rise to peak), which puts it in the realm of superoutburst amplitudes and makes it atypical for normal precursor outbursts. The amplitudes of the latter are in the range 1-2~mag. 
\item Its precursor decay of 3-5~mag is an outlier among the superoutbursts of DNe with a precursor. 
\end{itemize}

\citet{Kimura-2020} recently presented observations of the 2018 superoutburst of EG~Cnc, another well-known WZ~Sge system, which was first caught during an initial decay in its light curve of around 2~mag (with a similar decline-rate as 2020iko). There were no observations prior to the decay, which was deeper than its previous superoutburst in 1996-1997, whose first segment showed early superhumps. Otherwise, the 2018 event was identical to the 1996-1997 superoutburst. \citet{Kimura-2020} interpreted this superoutburst as similar to that of ordinary SU~UMa type systems. These authors argued that the 2018 superoutburst was associated with the extension of the accretion disk falling short of the 2:1 resonance because of a reduced amount of mass in the disk. EG~Cnc underwent a normal outburst in 2009, which could have depleted its disk mass. Hence, they considered the initial decay in the light curve to be a precursor to the superoutburst.

We speculate the current superoutburst of 2020iko may be similar to the 2015 superoutburst of AL~Com and 2018 superoutburst of EG~Cnc (provided the initial decay in its light curve was indeed a precursor), where the accretion disk expanded just up to the 3:1 resonance radius. Then, the large decay of 3-5~mag in 2020iko's precursor indicates a relatively slow growth of the tidal instability, which powers the main outburst. Direct evidence of tidal instability would have been provided by high-cadence (minute-scale) observations in this decay branch (as was the case for V1504~Cyg and V644~Lyr observed with {\it Kepler}), which unfortunately are not available for this source. The growth rate of this particular tidal instability is proportional to $q^{2}$, where $q$ is the mass-ratio of the system \citep{low-alpha}. WZ~Sge systems are characterized by very low values of $q$ (typically 0.07-0.08; \citealt{Kato-2015}). Thus, the large decay of 2020iko's precursor would imply that its mass-ratio is extremely small -- possibly smaller than that of EG~Cnc, for which $q\sim0.048$ \citep{Kimura-2020}. Detailed theoretical modeling of the multi-band light curves of 2020iko is beyond the scope of this paper, however, such work would help to provide insight into its extreme characteristics and throw light on why such precursors are rare in these low-$q$ systems.

\section{Conclusion} \label{sec:con}
We report the first result from a successful beta implementation of the anomaly detection algorithm of \citet{Soraisam-2020} in the real-time alert environment within ANTARES. 2020iko, located close to an extended source, was flagged from the ZTF public alert stream as a candidate peculiar source (see Fig.~\ref{fig:lc}). We undertook spectroscopic as well as photometric multi-wavelength follow-up observations, in particular, using LCO and MITSuME facilities. The fully assembled light curve of 2020iko reveals a precursor, a main superoutburst, and rebrightenings, while the spectrum taken during the superoutburst exhibits a blue continuum with higher-order Balmer lines in absorption, typical of DNe. The precursor has an amplitude of 6.5~mag and the superoutburst, 7~mag. Based on ASAS-SN data, the source, however, has no historical outburst going back to 2014, pointing to a recurrence period of $>6$~years. Taking together all these results, we conclude 2020iko to be a WZ~Sge-type DN candidate. Its precursor outburst with a large amplitude of 6.5~mag and a decay spanning $3\mbox{--}5$~mag is unprecedented for DNe. We speculate this event of 2020iko to be a superoutburst with a limited expansion of the accretion disk, falling within the 2:1 resonance radius reached in typical superoutbursts in other WZ~Sge systems. Detailed theoretical modeling of the data presented here will help to throw light on the factors leading to atypical limited expansion of accretion disks in superoutbursts in WZ~Sge type DNe.

\acknowledgments
We thank the referee for the helpful comments that have helped improve the paper. This work makes use of observations from the LCO network. MDS is supported by the Illinois Survey Science Fellowship of the Center for Astrophysical Surveys at the University of Illinois at Urbana-Champaign. DJS acknowledges support from NSF AST-1813466. This work was supported in part by Grant-in-Aid for Scientific Research on Innovative Areas (17H06362), Optical and Near-Infrared Astronomy Inter-University Cooperation Program and the joint research program of
the Institute for Cosmic Ray Research (ICRR).

\facilities{LCO, {\it Swift} (XRT and UVOT)}

\software{\texttt{numpy} \citep{numpy}, \texttt{scipy} \citep{scipy}, \texttt{astropy} \citep{astropy}, \texttt{matplotlib} \citep{matplotlib}, HEASOFT~\citep{heasoft}}

\begin{table*}
\caption{Light curve data}\label{tab:lc_data}
\renewcommand\arraystretch{1.0}
\renewcommand{\tabcolsep}{12pt}
\centering
\begin{tabularx}{0.6\textwidth}{llllc}
\hline
MJD    &Source          &Passband       &Mag    &Mag\_error\\
\hline
58965.200903   &ZTF   &$g$   &16.74   &0.03\\
58966.271000   &ATLAS   &$o$-ATLAS   &18.56   &0.09\\
58968.180278   &ZTF   &$r$   &19.80   &0.14\\
58968.216215   &ZTF   &$g$   &19.89   &0.14\\
58971.151817   &ZTF   &$r$   &16.46   &0.03\\
58971.180544   &ZTF   &$r$   &16.47   &0.03\\
58971.239954   &ZTF   &$g$   &16.18   &0.03\\
58972.166701   &ZTF   &$r$   &16.47   &0.03\\
58973.106501   &LCO   &$gp$   &16.18   &0.08\\
58973.106852   &LCO   &$gp$   &16.26   &0.08\\
58973.107196   &LCO   &$gp$   &16.16   &0.07\\
58973.107647   &LCO   &$rp$   &16.53   &0.11\\
58973.107990   &LCO   &$rp$   &16.56   &0.12\\
58973.108341   &LCO   &$rp$   &16.61   &0.12\\
58973.108756   &LCO   &$ip$   &16.62   &0.13\\
58973.109092   &LCO   &$ip$   &16.69   &0.14\\
58973.109435   &LCO   &$ip$   &16.91   &0.17\\
58973.109886   &LCO   &$up$   &16.82   &0.10\\
58973.110315   &LCO   &$up$   &16.62   &0.12\\
58973.110730   &LCO   &$up$   &16.90   &0.10\\
58973.111259   &LCO   &$zs$  &16.60   &0.14\\
58973.111681   &LCO   &$zs$   &16.86   &0.17\\
58973.112096   &LCO   &$zs$   &17.12   &0.20\\
58973.455033   &MITSuME   &$R_{c}$   &16.43   &0.14\\
58973.455033   &MITSuME   &$gp$   &16.18   &0.12\\
58973.455033   &MITSuME   &$I_{c}$   &16.67   &0.19\\
\hline
\end{tabularx}
\tablenotetext{}{(The full version is available in machine-readable form online.)}
\end{table*}

\bibliography{ref}{}
\bibliographystyle{aasjournal}

\end{document}